\documentclass{DISproc}
\usepackage{psfrag,epsfig,graphicx,graphics,xcolor}
\usepackage{wrapfig}

\newcommand{\xb}{{\underline x}}

\newcommand{\kb}{{\underline k}}

\newcommand{\be}{\begin{equation}}
\newcommand{\ee}{\end{equation}}
\newcommand{\eq}{\end{equation}}

\begin{document}
\title{High Energy exclusive Leptoproduction of the $\rho-$meson: Theory and Phenomenology}

\author{{\slshape I.~V.~Anikin$^{1,2}$, A. Besse$^3$, D.Yu.~Ivanov $^4$, B.~Pire$^5$, L. Szymanowski$^6$, S. Wallon$^{3,7}$}\\[1ex]
$^1$Bogoliubov Laboratory of Theoretical Physics, JINR, 141980 Dubna, Russia\\
$^2$Institute for Theoretical Physics, University of Regensburg, D-93040 Regensburg, Germany\\
$^3$LPT, Universit\'e Paris-Sud, CNRS, 91405, Orsay, France\\
$^4$Sobolev Institute of Mathematics and Novosibirsk State University, 630090 Novosibirsk, Russia\\
$^5$CPHT, {\'E}cole Polytechnique, CNRS, 91128 Palaiseau Cedex, France\\
$^6$National Center for Nuclear Research (NCBJ), Warsaw, Poland\\
$^7$UPMC Univ. Paris 06, facult\'e de physique, 4 place Jussieu, 75252 Paris Cedex 05,
France}

%
%
%
%
%
%
\contribID{xy}

\doi  

\maketitle

\begin{abstract}
 We describe the hard leptoproduction of transversally polarized rho-meson, up to twist~3 accuracy, including 2- and 3- particles Fock-states, in the HERA kinematics of high center-of-mass energy. We first build a model based on a simple approach to the unintegrated gluon density (the parton impact factor) that we compare with H1 and ZEUS data for the ratios of helicity amplitudes $T(\gamma^{*}_T \to \rho_T)/T(\gamma^{*}_L \to \rho_L)$ and $T(\gamma^{*}_T \to \rho_L)/T(\gamma^{*}_L \to \rho_L)$ and get a good description of the data. We also show how saturation effects can be included in this model by extending the dipole representation of the scattering amplitude in coordinate space up to twist~3. 
%
%
\end{abstract}

\section{Ratios of helicity amplitudes of the hard leptoproduction of the $\rho-$meson: a theoretical approach}
\begin{figure}[htb]
\centering
\psfrag{rho}[cc][cc]{$\hspace{.8cm}\rho (p_{\rho})$}
\psfrag{k}[cc][cc]{$k$}
\psfrag{rmk}[cc][cc]{}
\psfrag{l}[cc][cc]{}
\psfrag{q}[cc][cc]{}
\psfrag{lm}[cc][cc]{\raisebox{.6cm
}{$\quad \quad \, \, \, \Gamma \ \,\quad \Gamma$}}
\psfrag{H}[cc][cc]{
$ H_{q \bar{q}}$}
\psfrag{S}[cc][cc]{
$ S_{q \bar{q}}$}
\begin{tabular}{ccc}
\hspace{-1.5cm}\epsfig{file=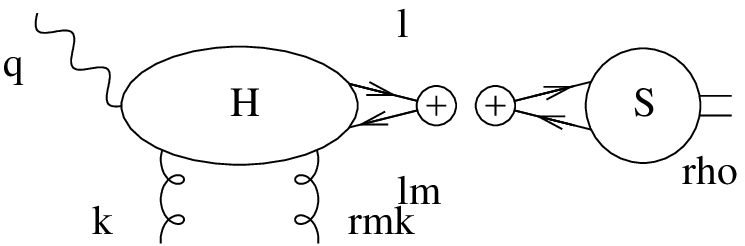,width=.45\linewidth}&\raisebox{4
    \totalheight}{+}&
\psfrag{H}[cc][cc]{
 $\partial_{\perp} H_{q \bar{q}}$}
\psfrag{S}[cc][cc]{
$ \partial_\perp S_{ q \bar{q}}$}
\epsfig{file=FiertzHSqq_rhofact.eps,width=.43\linewidth}\\
&\raisebox{3.5
    \totalheight}{\hspace{-.4\linewidth}+}
  &
    \psfrag{k}[cc][cc]{}
\psfrag{rmk}[cc][cc]{}
\psfrag{l}[cc][cc]{}
\psfrag{q}[cc][cc]{}
\psfrag{Hg}[cc][cc]{
$H_{q \bar{q}g}$}
\psfrag{Sg}[cc][cc]{
$\!S_{q \bar{q}g}$}
\psfrag{H}[cc][cc]{
$H_{q \bar{q}g}$}
\psfrag{S}[cc][cc]{
$ S_{q \bar{q}}$}
\psfrag{S}[cc][cc]{
$ \tilde{S}_{ q \bar{q}g}$}
\psfrag{lm}[cc][cc]
{\raisebox{.2cm
}{$\quad \,\,\,\, \, \, \, \Gamma \ \, \, \Gamma$}}
\hspace{-.4\linewidth}\epsfig{file=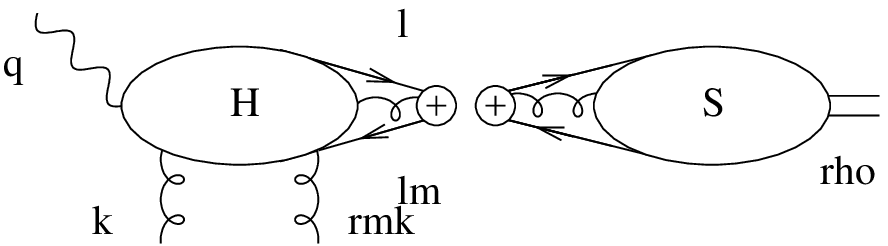,width=.45\linewidth}
\end{tabular}
\caption{Illustration of the 3 different types of terms entering the computation up to twist~3. From left to right: the 2-parton collinear contribution, the 2-parton first order contribution in $\ell_{\perp} $, the collinear term of the 3-parton contribution. $\Gamma$ stands for the set of Dirac matrices inserted using Fierz identity in the spinor space.}
\label{fig1}
\end{figure}

\begin{figure}[h!]
\begin{center}
\includegraphics[width=6cm]{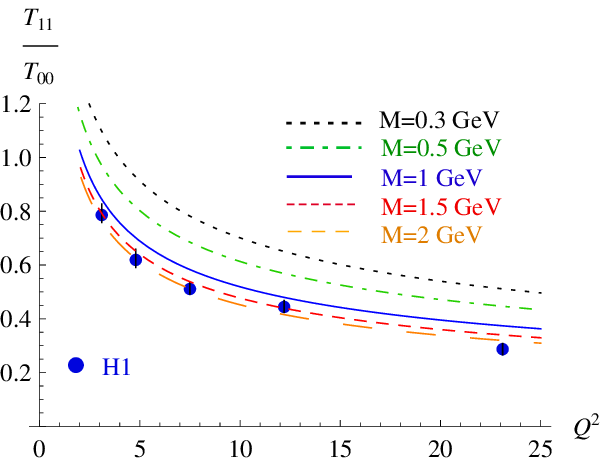} \includegraphics[width=6cm]{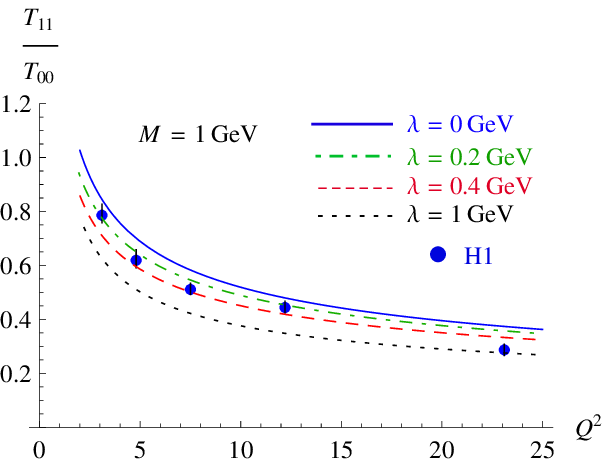}
\caption{Predictions for the ratio $T_{11}/T_{00}$ as a function of $Q^2$, compared to the data from H1~\cite{H1}.
Left: fixed $\lambda = 0$, and various values for $M$. Right: fixed scale M = 1\:GeV, and various values of $\lambda$.}
\label{fig1.2}
\end{center}
\end{figure}

In the Regge inspired factorization scheme, helicity amplitudes of the hard diffractive $\rho-$meson production 
\begin{displaymath}
\gamma^{*}(q,\lambda_{\gamma}) \, N(p_2) \rightarrow \rho (p_{\rho},\lambda_{\rho}) \, N(p_2)
\end{displaymath}
 are expressed in terms of the $\gamma^*(\lambda_{\gamma})\to \rho(\lambda_{\rho})$ impact factor ($\Phi^{\gamma^{*}(\lambda_{\gamma}) \rightarrow \rho(\lambda_{\rho})}$) and the nucleon transition impact factor ($\Phi^{N \rightarrow N}$).
At Born order, the helicity amplitudes read (using boldface letters for the euclideans two dimensional transverse vectors):
\begin{eqnarray}
\label{defImpactRep}
 T_{\lambda_{\rho}\lambda_{\gamma}}(\textbf{r};Q , M)= is\int \frac{d^2\textbf{k}}{(2\pi)^2}\frac{1}{\textbf{k}^2(\textbf{k}-\textbf{r})^2}\,\Phi^{N \rightarrow N} (\textbf{k},\textbf{r};M^2)\,\Phi^{\gamma^{*}(\lambda_{\gamma}) \rightarrow \rho(\lambda_{\rho})}(\textbf{k},\textbf{r};Q^2)\,.
 \end{eqnarray}
The momenta $q$ and $p_{\rho}$ are parameterized using the Sudakov decompositions in terms of two light cone vectors $p_1$ and $p_2$ as $q=p_1-\frac{Q^2}{s}p_2$ and $p_{\rho}\equiv p_1+\frac{m_\rho^2-t+t_{min}}{s}p_2+r_{\perp}$ with $2 p_1 \cdot p_2=s $ and $q^2=-Q^2$ is the virtuality of the photon.
The computation of the $\gamma^{*}(\lambda_{\gamma})  \rightarrow \rho (\lambda_{\rho}) $ impact factor is performed within collinear factorization of QCD. 
The dominant contribution (twist~2) $\gamma^{*}_L  \rightarrow \rho_L$ transition (twist~2) has been computed long time ago~\cite{GinzburgPanfilSerbo} while a consistent treatment of the twist~3 $ \gamma^{*}_T  \rightarrow \rho_T$ transition has been performed only recently in references~\cite{Anikin2009,Anikin2010}. 
Impact factors involve a hard part where the hard photon decays into partons that interact with the $t-$channel gluons and soft parts where these partons hadronize into a $\rho-$meson. Soft and hard parts are factorized in momentum space by expanding the hard parts around the longitudinal components of the momenta of the partons
, collinear to the direction of the $\rho-$meson momentum. Fierz identity applied to spinor and color spaces, is used to factorize color and spinor indices linking hard and soft parts. 
Up to twist~3, the amplitude involves the contributions of the 2- and 3-parton exchanges between the hard part and the soft part of the impact factor, Figure~\ref{fig1} shows the 3 types of terms involved.
Soft parts correspond to the Fourier transforms of the vacuum to $\rho-$meson matrix elements that are parameterized by five 2-parton distribution amplitudes (DAs) and two 3-parton DAs. Due to the equation of motion of QCD and the so-called condition of the $n-$independence (c.f~\cite{Anikin2010} for details), the DAs are expressed in terms of a single pure 2-parton DA and two pure 3-parton DAs. Results can be then split into the 2-parton contribution, so-called Wandzura-Wilczek (WW) contributions, and the genuine 3-parton contribution. In our model we take the explicit solutions of the ERBL equations up to the first term involving the factorization scale dependence $\mu_F$.
 A first phenomenological approach~\cite{Anikin2011} consists in choosing a simple model for the proton impact factor~\cite{GunionSoper}, depending on two free parameters $M$ and $A$,
\begin{displaymath}
\label{ProtonIF}
\Phi_{N \to N}(\textbf{k},\textbf{r};M^2)\!\!=\!\! A \delta_{ab}\!\!\left[\frac{1}{M^2+(\frac{\textbf{r}}{2})^2}-\frac{1}{M^2+(\textbf{k}-\frac{\textbf{r}}{2})^2}\right]\!\!.
\end{displaymath}
Figure~\ref{fig1.2} shows the results for the ratios of the helicity amplitudes that we obtain with this approach, compared to the data of H1~\cite{H1} depending on the free parameter $M$ and on an infra-red cut off $\lambda$ for the integral over the transverse momenta of the $t-$channel gluons. Similarly, ZEUS~\cite{ZEUS} data were compared to the results. Predictions are in fairly good agreement with the data for a value of $M$ between 0.9\:GeV and 5\:GeV and the result depends weakly of the value of $M$ and $\mu_F$. The dependence in the cut-off $\lambda$ shows that soft gluons with momenta smaller than $\Lambda_{QCD}$, $\left|\textbf{k}\right|\leq \Lambda_{QCD}$, have a small contribution to the result while the contribution of the gluons with $\left|\textbf{k}\right|\sim $1\:GeV cannot be neglected and thus calls for an inclusion of the saturation dynamics of the nucleon.
\section{Dipole representation of the scattering amplitude in coordinate space up to twist~3}
In the Ref.~\cite{Besse2012}, we show that performing the collinear approximation up to twist~3 after expressing the hard and soft parts of the impact factor $\gamma^*_T\to \rho_T$ in term of their transverse Fourier transforms, the color dipole interaction with the $t-$channel gluons factorizes out in the scattering amplitude of the full process. 
In the WW approximation, the result exhibits the wave function of the tranversally polarized photon $ \Psi_{\lambda{\gamma} (\lambda)}^{\gamma^*_{T}}$~\cite{Mueller,NikolaevZakharov,IvanovWusthoff,GiesekeQiao} and the relevant combination of DAs $\phi_{\lambda_{\rho} (\lambda)}^{WW}$,
\begin{align}
\label{A-2parton-final-non-flip}
\Phi^{\gamma^*_T \to \rho_T,\, WW}= \frac{m_{\rho}\, f_{\rho}}{\sqrt{2}}\int dy\ \int d^2\xb\, g^2 \,\delta^{ab}\, \mathcal{N}(\xb,\kb)\,\sum_{(\lambda)}  \phi_{\lambda_{\rho} (\lambda)}^{WW}
\, \Psi_{\lambda{\gamma} (\lambda)}^{\gamma^*_{T}} \,,
\end{align}
where $\lambda$ denotes the helicity of the exchanged quark. In Eq.~\ref{A-2parton-final-non-flip}, $\mathcal{N}(\xb,\kb)$ is proportional to the scattering amplitude of a color dipole with the two $t-$channel gluons.
\begin{figure}[h!]
\psfrag{gam}[cc][cc]{ $\Psi_{\lambda{\gamma} (\lambda)}^{\gamma^*_{T}}$}
\psfrag{fl}[cc][cc]{$\xb$}
\psfrag{hq}[cc][cc]{$\lambda$}
\psfrag{X}[cc][cc]{$\!\times \quad \xb \ $}
\psfrag{hqb}[cc][cc]{} 
\psfrag{Soft}[cc][cc]{$\quad \phi_{\lambda_{\rho} (\lambda)}^{WW}$}
\psfrag{hqbs}[cc][cc]{}
\psfrag{hqs}[cc][cc]{$\lambda$}
\psfrag{gami}[cc][cc]{{\footnotesize $\Gamma$}}
\centerline{\epsfig{file=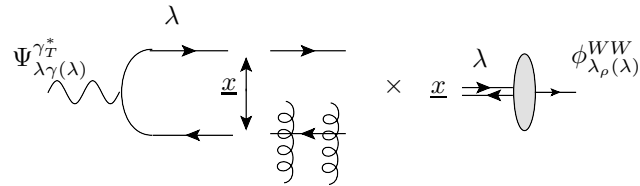,width=8cm}}
\caption{Illustration of the $\gamma^*_T\to\rho_T$ in the WW approximation in the dipole factorized form after using the equation of motion of QCD.}
\label{Fig:2-parton-dipole}
\end{figure}
Let us emphazise that the appearance in Eq.~\ref{A-2parton-final-non-flip} of the scattering amplitude of the dipole with the $t-$channel gluons $\mathcal{N}(\xb,\kb)$ is not a straightforward consequence of the Fourier transform of the hard part. The factorized form in  Eq.~\ref{A-2parton-final-non-flip} appears only after using the equation of motion of QCD.

Beyond the WW approximation, the 3-parton contribution leads to a similar result where the dipole scattering amplitude factorizes out after using the equation of motion of QCD. However, since the 3-body wave function of the photon is still unknown, a clear interpretation of this factorized form is still required.
%
\section{Conclusion}
We have first shown~\cite{Anikin2011} that a model \`a la Gunion and Soper gives a good description of the data, however saturation effects are needed. We then proved in Ref.~\cite{Besse2012} that the twist~3 impact factor of the transition $\gamma^*_T\to\rho_T$  
 exhibits the color dipole scattering amplitude with the target. The next step is now to compare H1 and ZEUS data with the ratio $T_{11}/T_{00}$ obtained by combining our results of the impact factor up to twist~3 computed in the collinear approximation with a model 
 for the color dipole/target amplitude. Another perspective would be to extend our computations to the nonforward kinematics, in particular in order to investigate saturation effects at fixed impact parameter. 
%

\section{Acknowledgements}

We thank K. Golec-Biernat and L. Motyka for stimulating discussions. This work is supported in part by the Polish NCN grant DEC-2011/01/B/ST2/03915, by the French-Polish Collaboration Agreement Polonium, by the Russian grant RFBR-11-02-00242 and by the P2IO consortium.

%
%
 

{\raggedright
\begin{footnotesize}



\end{footnotesize}
}


\end{document}